\documentclass[iop,superscriptaddress]{emulateapj}
\usepackage{amsmath,epsfig,url,natbib}
\usepackage{graphicx,epstopdf,float,color,array}
\usepackage{rotating,multirow}

\shorttitle{[CII] Line Luminosity Function at $z\sim0$}
\shortauthors{Hemmati et al.}

\newcommand{\iras}{{\sl IRAS}}

\newcommand{\herschel}{{\sl Herschel}}

\newcommand{\alma}{{\sl ALMA}}

\newcommand{\degree}{\ensuremath{^\circ}}
\newcommand{\CII}{[C\,{\sc ii}]}
\newcommand{\NII}{[N\,{\sc ii}]}

\begin{document}
\title{The Local [CII]\,158$\mu$\lowercase{m} Emission Line Luminosity Function}
\author{
Shoubaneh Hemmati\altaffilmark{1},
Lin Yan\altaffilmark{1},
Tanio Diaz-Santos\altaffilmark{3}, 
Lee Armus\altaffilmark{2},
Peter Capak\altaffilmark{1,2},
Andreas Faisst\altaffilmark{1},
Daniel Masters\altaffilmark{1}}

\email{shemmati@ipac.caltech.edu}
\altaffiltext{1}{Infrared Processing and Analysis Center, Department
  of Astronomy, California Institute of Technology, 1200 E. California Blvd., Pasadena CA 91125, USA}
\altaffiltext{3}{Nucleo de Astronomia de la Facultad de Ingenieria, Universidad Diego Portales, Av. Ejercito Libertador 441, Santiago, Chile}
\altaffiltext{2}{Spitzer Science Center, Department of Astronomy,
 California Institute of Technology, 1200 E. California Blvd,
  Pasadena, CA 91125, USA}

\begin{abstract}

 We present, for the first time, the local \CII\ 158 $\mu$m emission line
 luminosity function measured using a sample of more than 500
 galaxies from the Revised Bright Galaxy Sample (RBGS). \CII\ luminosities are measured from the \herschel\ PACS
 observations of the Luminous Infrared Galaxies in the Great
 Observatories All-sky LIRG Survey (GOALS) and estimated for
 the rest of the sample based on the far-IR luminosity and color.  The sample covers 91.3\% of the sky and is
 complete at $S_{60\mu m} > 5.24\  \rm Jy$. We calculated the completeness as a function of
 \CII\ line luminosity and distance, based on the far-IR color and
 flux densities. The \CII\ luminosity function is constrained in the range
 $\sim 10^{7-9} \ L_{\odot}$ from both the 1/V$_{max}$ and a maximum likelihood methods. The shape of our derived \CII\ emission line luminosity function agrees well with the IR luminosity function. For the CO(1-0) and \CII\ luminosity functions to agree, we propose a varying ratio of \CII/CO(1-0) as a function of CO
 luminosity, with larger ratios for fainter CO luminosities. Limited
 \CII\ high redshift observations as well as estimates based on the IR and UV luminosity
 functions are suggestive of an evolution in the \CII\ luminosity function
 similar to the evolution trend of the cosmic star formation rate
 density. Deep surveys using ALMA with full capability will be able to confirm this
 prediction.

\end{abstract}


\section{Introduction}

The gas content in the interstellar medium (ISM) of galaxies is critical in galaxy evolution,
serving as the immediate fuel for star formation
(\citealt{Scoville2016}). Rotational transitions of common
interstellar molecules, such as CO, as well as atomic fine structure line transitions,
predominantly \CII, can be used to study the amount and distribution
of the cold gas content in galaxies (e.g. \citealt{Carilli2013}). 

The \CII\ fine structure line at 157.74 $\mu$m which arises from
the transition of singly ionized carbon atoms (C$^+$) from $^2P_{3/2}$ to
$^2P_{1/2}$ state, is the strongest emission line in the
far-infrared (FIR). The primary mechanism for producing this line is
excitation of C$^+$ atoms via collisions with other particles such as
neutral hydrogen (H) or free electrons and protons
(e.g. \citealt{Hayes1984}). The ionization potential of C$^+$ is quite
shallow, only 11.26\,eV, and the critical density with collisions with
neutral and molecular hydrogen $n^{cr}_{H}$ is also small, for $T=100k$
  $\sim 3\times10^3$\,cm$^{-3}$ (\citealt{Goldsmith2012}).
Therefore, the 158$\mu$m line is an efficient and dominating coolant
for neutral gas. For nearby normal star forming galaxies as well as
Luminous Infrared Galaxies (LIRG), the 158$\mu$m line, in combination with far-IR
continuum, CO (1-0) and \NII , provides powerful spectral diagnostics
of physical state of the ISM, such as intensity of FUV radiation field,
temperature, density and chemical abundance
(e.g. \citealt{malhotra1997, kaufman1999, Stacey2010,
  Nagao2012}). \CII\ emission can be produced in both neutral as well as ionized phases of
the interstellar medium (ISM). For example, in the Milky Way,
  \cite{Pineda2014} measured the contribution of the ionized phase of
  the ISM to \CII\ luminosity to be around 20\% and the remaining 80\% coming from the neutral gas.\citealt{Goldsmith2015} showed that the
  contribution from the ionized region can be as high as 50\%   using
  PACS observations of ionized nitrogen in the galactic plane. In
other galaxies, the fraction of \CII\ arising from the ionized regions
compared to neutral parts is more uncertain and still a matter of
study. It has been shown to vary for different galaxies and to depend
on the properties of the ISM (e.g. \citealt{Cormier2012}, \citealt{Decarli2014},
\citealt{Gullberg2015}, \citealt{Olsen2015}). 

Many studies have focused on the \CII\ emission line
luminosity as a star formation rate indicator, as it is a very bright line almost
unaffected by extinction (e.g. \citealt{Stacey1991},
\citealt{Boselli2002}, \citealt{Gracia2011}, \citealt{Sargsyan2012}, \citealt{Delooze2014},
\citealt{Vallini2015}, \citealt{Brisbin2015}). The major limitation for using
the \CII\ luminosity to measure the star formation rate is the
so-called \CII\ deficit, which corresponds to lower fraction of \CII\
to far-IR as a function of increasing warm infrared color
(e.g. \citealt{malhotra1997}, \citealt{Malhotra2001},
\citealt{Santos2013}). More recently, \cite{Santos2014} using a sample
of luminous local LIRGs found that the \CII\ deficits are restricted
to their nuclei. \cite{Herrera2015} using the resolved \CII\
observations of \herschel\ KINGFISH (\citealt{Kennicutt2011}) galaxies
also showed that the \CII\ surface density correlates well with
star formation rate surface density both globally and in kpc-scale in
the absence of strong active galactic nuclei.

During the next few years, with the steadily improved sensitivity and
frequency coverage, the Atacama Large Millimeter Array (\alma) will
dramatically increase the number of galaxies with detected \CII\
emission at high redshift (e.g. \citealt{Swinbank2012},
  \citealt{Capak2015}, \citealt{Aravana2016}), making systematic surveys possible.  One
powerful application of \CII\ , the brightest of far-IR emission lines,
is for measuring redshifts of distant ($z\ge6$) galaxies in the early
Universe.  Similar to commonly seen galaxy redshift surveys based on
optical spectroscopy,  these \alma\ \CII\ surveys will characterize
the abundance and intensity distributions of \CII\ emitters by
deriving the line luminosity functions. Future \CII\ redshift surveys would require a well-measured line
luminosity function at $z\sim0$ for comparison. 

 Previous studies have briefly looked at the \CII\ line local
  luminosity function using either FIR luminosity functions or limited
  luminosity range observed data with complex selection functions
  (e.g. \citealt{Swinbank2012}, \citealt{Brauher2008}). The goal of
  this paper is to obtain the $z\sim0$ \CII\ luminosity function bench
mark. \herschel\ space observatory (\citealt{Pilbratt2010}), with its sensitive
far-IR spectroscopy and fast survey speed, has produced large samples
of galaxies with \CII\ detections at various redshifts
(e.g. \citealt{Santos2013,Farrah2013,Herrera2015}). In the local Universe, the far-IR spectra of  a complete
sample of Luminous/Ultraluminous Infrared Galaxies (LIRGs/ULIRGs) from
the Great Observatories All-sky LIRG Survey (GOALs)
(\citealt{Armus2009,Santos2013}) is the primary dataset we use for our analysis because it is a complete set of \CII\ observations of the Revised Bright Galaxy Sample (RBGS - Sanders et al .2003). In \S 2 we discuss in detail the selection of the local
sample and its completeness.  We present the local \CII\ line luminosity
function in \S 3.  We discuss and compare our results to other
indirect methods of estimating the \CII\ line luminosity function in
the local universe and also predicts the evolution of \CII\ line luminosity function
from existing UV observations in \S 4. The summary of the paper is
presented in \S 5. Throughout the paper, we adopt a flat concordance $\Lambda$CDM
cosmology, with $\Omega_{m}=0.28$, $\Omega_{\Lambda}=0.72$, and
$H_{0}$=70~km~s$^{-1}$~Mpc$^{-1}$. 

\section{Sample \label{sec:data}}

\begin{figure*} []
\centering
\begin{tabular}{cc}
  \includegraphics[trim=1.5cm 12.5cm 1.5cm 0cm, clip,
width=0.5\textwidth]{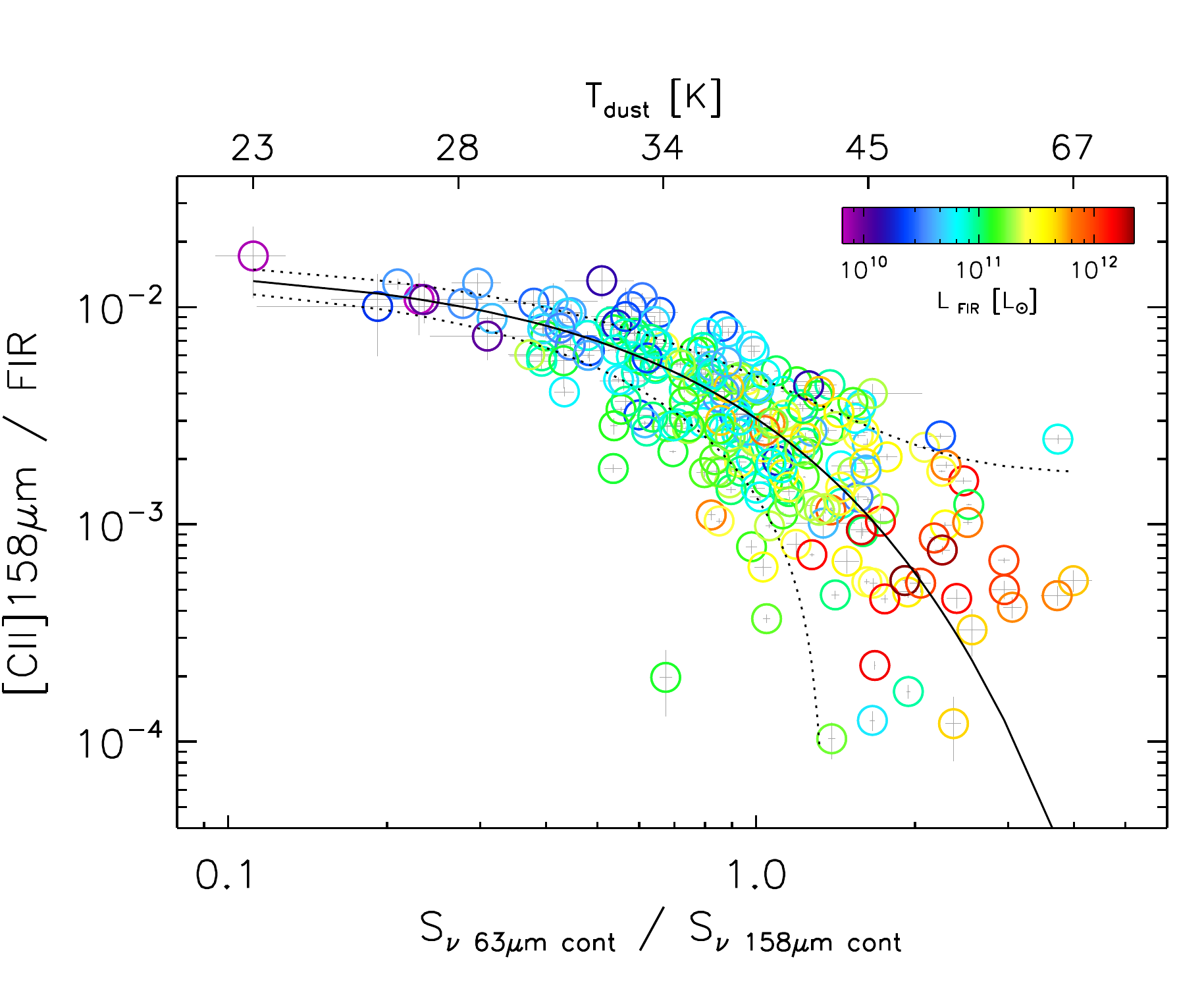}
\includegraphics[trim=1.5cm 12.5cm 1.5cm 0cm, clip,
  width=0.5\textwidth]{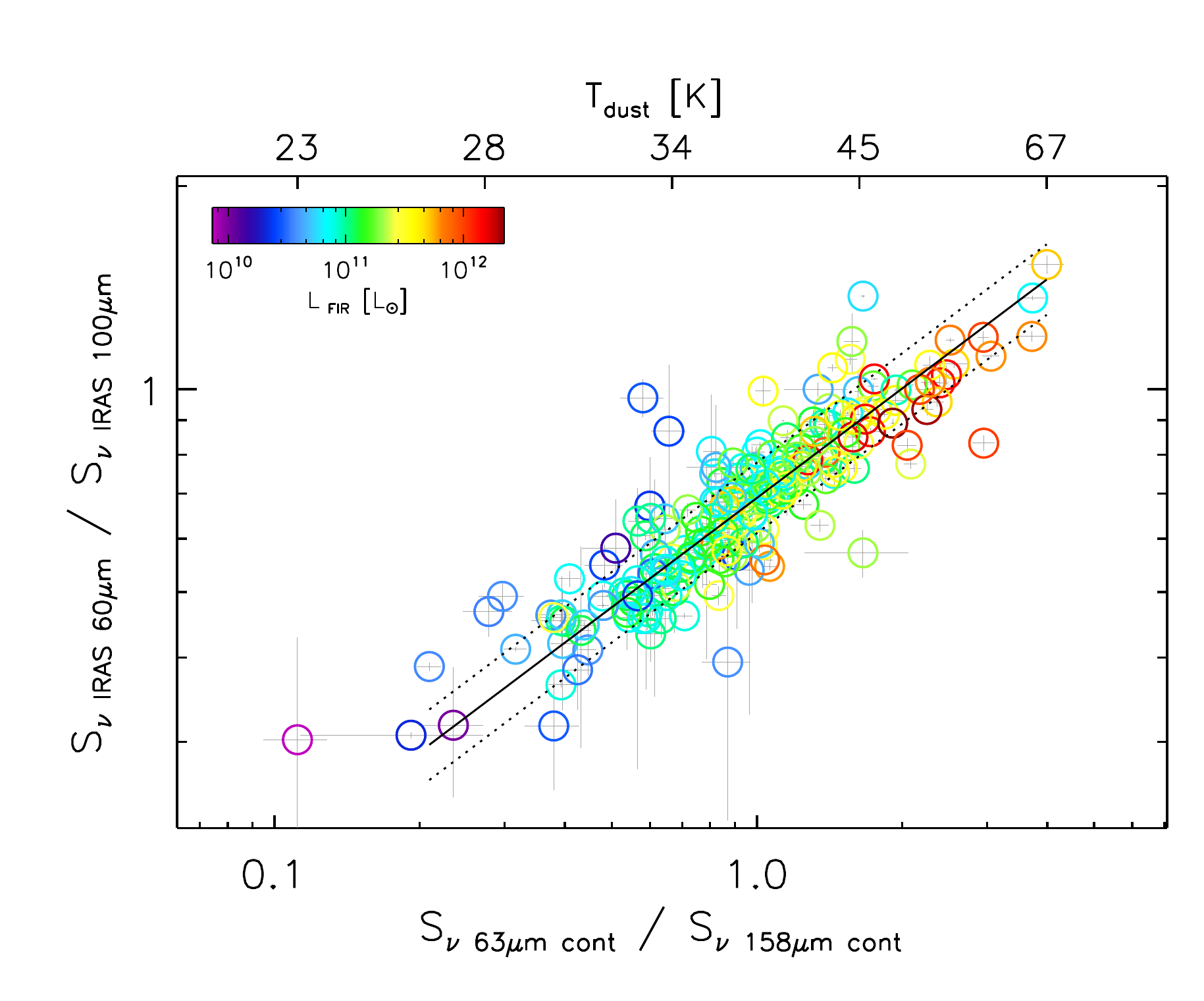}\\
\end{tabular}
\caption{(Left)  $f$(\CII)/$f$(FIR) versus
far-IR color $S_{63\mu m}/S_{158\mu m}$  of the galaxies in the GOALS
sample color-coded by the IR luminosity and (right) \iras\ -based far-IR color $S_{60\mu m}/S_{100\mu m}$ vs. the
  PACS based color $S_{63\mu m}/S_{158\mu m}$. The solid and dashed black
lines corresponding to the best fitted relation and 1$\sigma$ dispersion are used to predict the \CII\ line
luminosity for the rest of the RBGS sample.}
\label{fig:sample}
\end{figure*}

The most ideal survey for [CII] line luminosity function is a blind
spectroscopic survey with a uniform flux sensitivity, covering a well
defined area of sky.  In reality, such a survey in far-IR spectroscopy
over a large area is not possible, especially at $z\sim0$.  The next
best available option is to utilize the [CII] observations of a
complete sample of local galaxies with a well defined selection
function. The RBGS sample contains a total of 629 galaxies and is a complete 60$\mu$m flux limited
sample of all galaxies satisfying the following criteria: (1) \iras\
flux density $S_{60\mu m}$\,$>$\,5.24\,Jy; (2) Galactic latitudes $\rm
| b |$\,$>$\,5\degree. 
The Great Observatories All-sky LIRG Survey (GOALS;
\citealt{Armus2009}) contains all (202) luminous infrared galaxies
$\rm L_{8-1000\mu m} \ge 10^{11} L_{\odot}$ in the RBGS. All GOALS sources have complete far infrared photometric and spectroscopic coverage from IRAS, Spitzer and Herschel.

The GOALS sample was observed by the Integral Field Spectrometer (IFS)
of the PACS instrument (\citealt{Poglitsch2010}) on board of \herschel.  For this paper, we took
the best \CII\,158$\mu$m line measurement from a variety  of
apertures depending on each individual source (Diaz-Santos et al., in
prep.). In short, the best measurement is chosen based on visual inspection of
each individual source. In most cases we have used the total FOV
($5\times 5$ spaxel) quantities, in few cases where two components were
resolved within the FOV, the centeral spaxel (corrected for aperture)
is used instead. \citet{Santos2013} and \citet{Santos2014} give the detailed description on how the data
was reduced and the \CII\,158$\mu$m line fluxes were measured. In
short, the \herschel\ Interactive Processing Environment (HIPE;
ver. 8.0) application was used to retrieve and process the
spectra. The \CII\ flux is then measured by integrating the
continuum-subtracted spectrum within the $\pm 3\sigma$ region around
the central position of the line, and the associated uncertainty is
calculated as the standard deviation of the continuum integrated over
the same range of the line. The
major portion of this far-IR spectroscopic dataset is from the program (\herschel\  OT1\_larmus), and the remaining is from
the public archive data, collected by three other programs
(KPGT\_esturm\_1, PI: E. Sturm; KPOT\_pvanderw\_1, PI: P. van der
Werf; OT1\_dweedman\_1, PI: D. Weedman).  Of the 202 (U)LIRGs from the
GOALS sample, 200 have \CII\ observations (IRASF08339+6517 and
IRASF09111-1007 have no PACS spectra).

\subsection{\CII\ Luminosities and Uncertainties}

Ideally, constraining the \CII\ line luminosity function down to faint
luminosities ($\sim 10^{(7-8)} L_{\odot}$), would require
far-IR spectroscopy of a complete sample of low IR luminosity
galaxies. However, such a dataset currently does not exist.  The
alternative best approach is to predict the \CII\ luminosities for
RBGS galaxies fainter than LIRGS in the GOALS sample ($\rm L_{IR} <
10^{11 }L_{\odot} $). The \CII\ luminosities are calculated using the
established correlation between $f$(\CII)/$f$(FIR) versus far-IR color, {\it e.g.} dust temperature ($T_{dust}$) (see
\citet{Santos2013} and references there in). This relation, as shown on
the left panel of Figure \ref{fig:sample}, is expressed as:\\

 \begin{equation}\label{eq:c263}
\frac{[CII]}{FIR} = 0.016(\pm 0.001) \times exp(\frac{\frac{S_{63\mu
  m}}{S_{158\mu m}}}{0.60 (\pm 0.038)})
\end{equation}

with a dispersion of 0.0017 dex. We used a modified black body
  function with an emissivity index of $ \beta =1.8$ and reference
  wavelength of 100$\mu m$, that reproduces the observed
  $S_{63\mu m}/S_{158\mu m}$ color with the dust temperature shown in the upper
  axis of Figure \ref{fig:sample}. The FIR fluxes covering the
40-500$\mu m$ are calculated
as: $FIR= 1.26\times 10^{-14}(2.58 S_{60\mu m} + S_{100\mu m}) \rm [W
m^{-2}]$ (\citealt{Helou2000}) where $S_{\nu}$ are in $\rm  [Jy]$ . To measure the far-IR color on the right hand side of equation (1),
we use the correlation between the PACS-based far-IR color $S_{63\mu
  m}/S_{158\mu m}$ and the more commonly used \iras\ -based color $S_{60\mu
  m}/S_{100\mu m}$. As expected and shown on the right panel of Figure
\ref{fig:sample}, the two colors correlate well.

 \begin{equation}\label{eq:6063}
Log(\frac{S_{60\mu m}}{S_{100\mu m}}) = -0.161(\pm 0.004) +0.539(\pm 0.018) Log (\frac{S_{63\mu
  m}}{S_{158\mu m}})
\end{equation}

with a dispersion of 0.052 dex. 

The 1$\sigma$ uncertainties of \CII\ luminosities for GOALS galaxies
are taken from \citep{Santos2014}. For the non-GOALS RBGS galaxies,
however, all the above-mentioned assumptions need to be accounted
for. We calculate these errors using a Boot strapping technique. We
perturbed $S_{60\mu m}$, $S_{100\mu m}$ flux densities far-IR fluxes
as well as the fitting parameters (from equations \ref{eq:c263} \& \ref{eq:6063}) by randomly drawing
values from normal distributions with their corresponding standard errors as width of the distributions to measure
the \CII\ line luminosity standard error. These 1$\sigma$ measured
uncertainties are largely dominated by the uncertainties of the fitted
parameters in Equation (\ref{eq:c263}) and (\ref{eq:6063}) and are an upper limit to the true error as the
uncertainty in the fitting parameters already account for
uncertainties in fluxes and flux densities. Our method allows us to quantify the uncertainties in the derived \CII\ luminosities for
non-GOALs galaxies. Fundamentally, these uncertainties are driven by
the intrinsic variations in \CII-to-FIR flux ratios. \CII-to-FIR ratios are affected by several
physical conditions, such as FIR temperature, neutral gas density, or
equivalently surface brightness (e.g. \citealt{Lutz2016}). \CII\ emission is collisionally
excited and can be suppressed in very high density regions. Another
possible physical process leading to \CII\ deficit is the reduction in the photoelectric heating efficiency due to the
  charging of the dust grains. Therefore, the scattering in the observed \CII/FIR versus $S_{63\mu
  m}/S_{158\mu m}$ relation reflects the physical diversity of ISM in different
galaxies. 

After excluding the very nearby galaxies (luminosity distances less than 1 Mpc) as well as those with predicted
\CII\ line luminosity less than the lowest PACS observed \CII\ line
luminosity ($10^{6.73} L_{\odot}$) from this sample, we are left with
200 GOALS and 395 RBGS non-GOALS galaxies spanning the redshift range of $0.00023-0.076$ and
 \CII \ line luminosities in the range  $10^{6.73-9.33} L_{\odot}$

\subsection{Completeness}

\begin{figure*}[]
\includegraphics[trim=0cm 0cm 0cm 0cm, clip, width=1.0\textwidth]{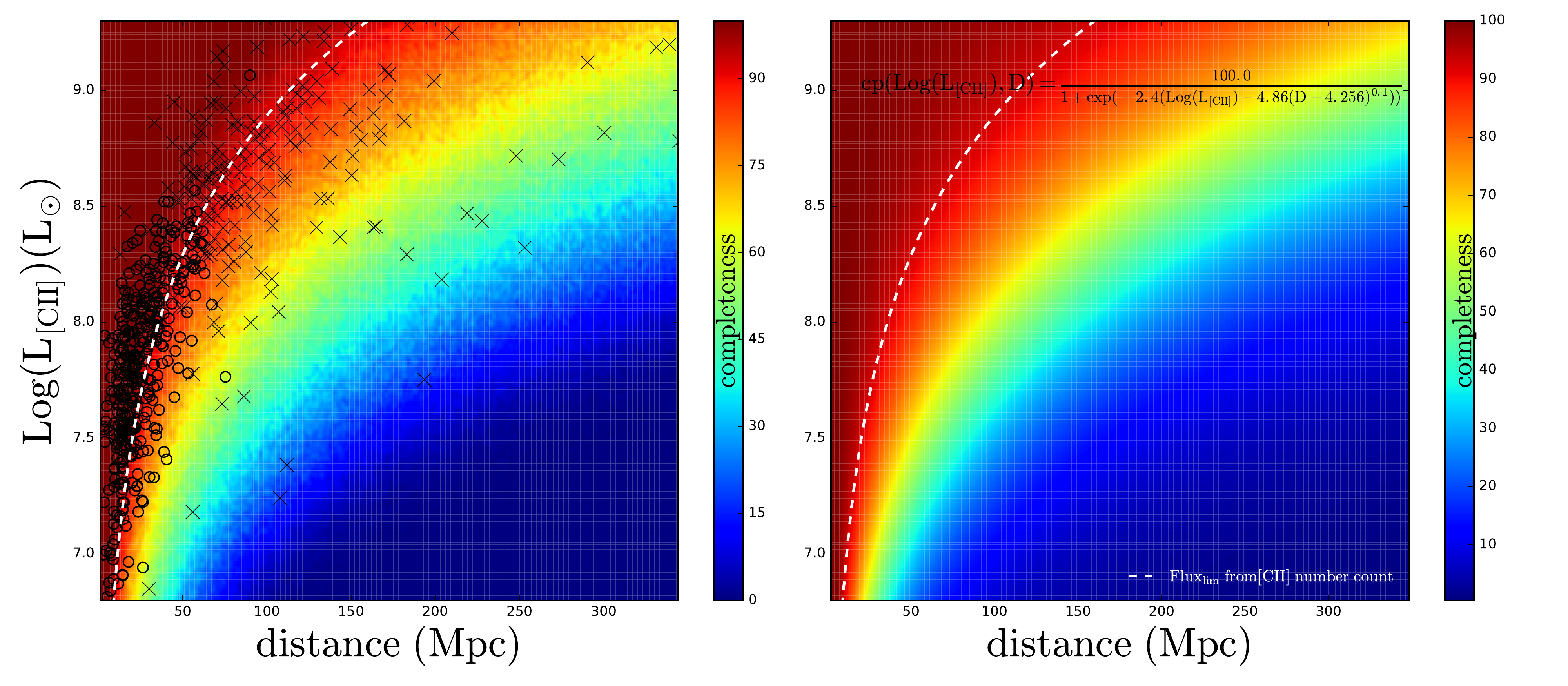}
\caption{Completeness percentage at each \CII \ line luminosity and
  distance (Left) measured numerical values from simulation. (Right)
  Analytical fitted function. The white dashed line which shows the 100 percent
  completeness as measured from the \CII \ flux number count turn
  over agrees very well with this method. Black crosses are our
  sources from the GOALS survey and black circles are the rest of
  the sample from the RBGS. 
}
\label{fig:completeness}
\end{figure*}

The sample we have selected here is complete at flux density $S_{60\mu
  m}$\,$>$\,5.24\,Jy. However, to estimate the \CII\ line luminosity
function, we need to know how incomplete the sample is at each \CII\
line flux. We therefore calculate a completeness function at each \CII\ line
flux ($C([CII],D_{L})$). This is critical for determining the faint end of the line luminosity
function. To estimate the amount of completeness at each \CII \ flux,
we make a grid of \CII \ line luminosity and distance. At each cell of
this grid we randomly draw a hundred far-IR colors ($S_{60\mu
  m}/S_{100\mu m}$) and use these ratios along with equations \ref{eq:c263} and
\ref{eq:6063} and their dispersion as well as the definition of FIR to
calculate the corresponding $S_{60\mu m}$. Then the completeness fraction at each
cell of the grid is simply the ratio of galaxies with $S_{60\mu
  m}$\,$>$\,5.24\,Jy (the selection criteria for the RBGS) to all hundred galaxies in that grid. The completeness function is shown
in the left panel of Figure \ref{fig:completeness}. As we are going to
use this completeness function in the luminosity function, to speed up
the calculation we fit an analytical form to this completeness
function. The best fitted function as shown on the right panel of
Figure \ref{fig:completeness} is expressed with a Sigmoid function as:
 
\begin{equation}\label{eq:compl}
\begin{aligned}
C([CII],D_{L})=\\
\frac{100}{1+exp(-2.4(Log([CII]) - 4.86(D_{L}-4.256)^{0.2}))}
\end{aligned}
\end{equation}

where the unit of \CII\ luminosities is $L_{\odot}$ and distances
are in Mpc. Completeness for different surveys are often just measured
by the turn over in the source counts as a function of
brightness. Here, we also show the \CII\ line flux at which the
source count starts to drop with white dashed line on Figure \ref{fig:completeness}
which agrees well with where our completeness function starts to drop.

In Figure \ref{fig:completeness} we also show where our galaxies sit
with black  crosses (GOALS objects) and plus symbols (the rest of the
RBGS) on top of the rainbow colored completeness values. The galaxies
are not originally selected based on their \CII\ luminosity and the
majority of the sample sits above 80\% completeness. We note that the
handful of galaxies with completeness values below 50\% are all from
the GOALS sample with observed \CII\ luminosities. The completeness
values will be used as weights for probabilities of individual sources in the sample. We explain the
details of implementing the completeness in the next section.

In addition to the source detection incompleteness, the sky coverage percentage is also
accounted for in our calculation. This galaxy sample spans over the entire sky except
for a thin strip within galactic latitude of $| b | < 5$\degree.  The
effective sky coverage is 37,657 square degrees, 91.3\%\ of the full
sky \citep{Sanders2003}. We include this multiplicative factor in the
luminosity function estimation.

\section{Luminosity Function}

Various methods exist for estimating the luminosity function. Each has
its own advantages and disadvantages. The most widely used
method is the 1/V$_{max}$ method (\citealt{Schmidt1968}, \citealt{Felten1976}) which assumes no parametric form for
the luminosity function and is very easy to implement. However, in
this method galaxies are binned into different luminosity bins and the choice of the bin size and centers of the
bins might affect the overall shape of the luminosity function. Another well-established method
is the Maximum likelihood estimator (\citealt{Sandage1979}) which has the
great advantage of using unbinned data. But unlike the 1/V$_{max}$ method, here a parametric form for the luminosity
function needs to be assumed. Also, the 1/V$_{max}$ method is only
accurate if there is a uniform density distribution. For large
enough surveys such as ours, this is not an issue. Here, we determine
the \CII\ line luminosity function using both of these methods.

\subsection{1/V$_{max}$}

We start with the 1/V$_{max}$ method to estimate the \CII\ line
luminosity function. This method was first discussed by
\citet{Schmidt1968} and revised later by \citet{Felten1976}.  With this
method, when we calculate volume number density of galaxies per $\rm
\Delta (L)$ for a flux limited sample, the relevant quantity is the
maximum volume (V$_{max}$) within which a galaxy could lie and still
be detected by the survey.  The underlying concept is that a brighter
galaxy can be seen further away than intrinsically fainter one, thus
probing a larger volume. This maximum volume is usually constraint by
both the maximum and minimum redshift an object could have and still
be included in the survey sample.

\begin{table*}
\centering
\caption{[CII] Luminosity Function Parameters}
\begin{tabular}{*{5}{c}}
\hline
\hline
Method & $\alpha$ & $\beta$ & $ L^{*} \rm (L_{\odot}) \times 10^{8}$ &
                                                                       $\phi \rm
                                                                 ^{*}(Mpc^{-3}
                                                                 log_{10}(L_{[CII]})^{-1})$ \\
\hline
\\
Maximum Likelihood&2.36$\pm$0.25&0.42$\pm$0.09&2.173$\pm$0.743&0.003$\pm$0.002\\
\\
\hline
\end{tabular}
\label{tbl:LF_param}
\end{table*}

\begin{figure*}[]
\includegraphics[trim=0cm 0cm 0cm 0cm, clip, width=1.02\textwidth]{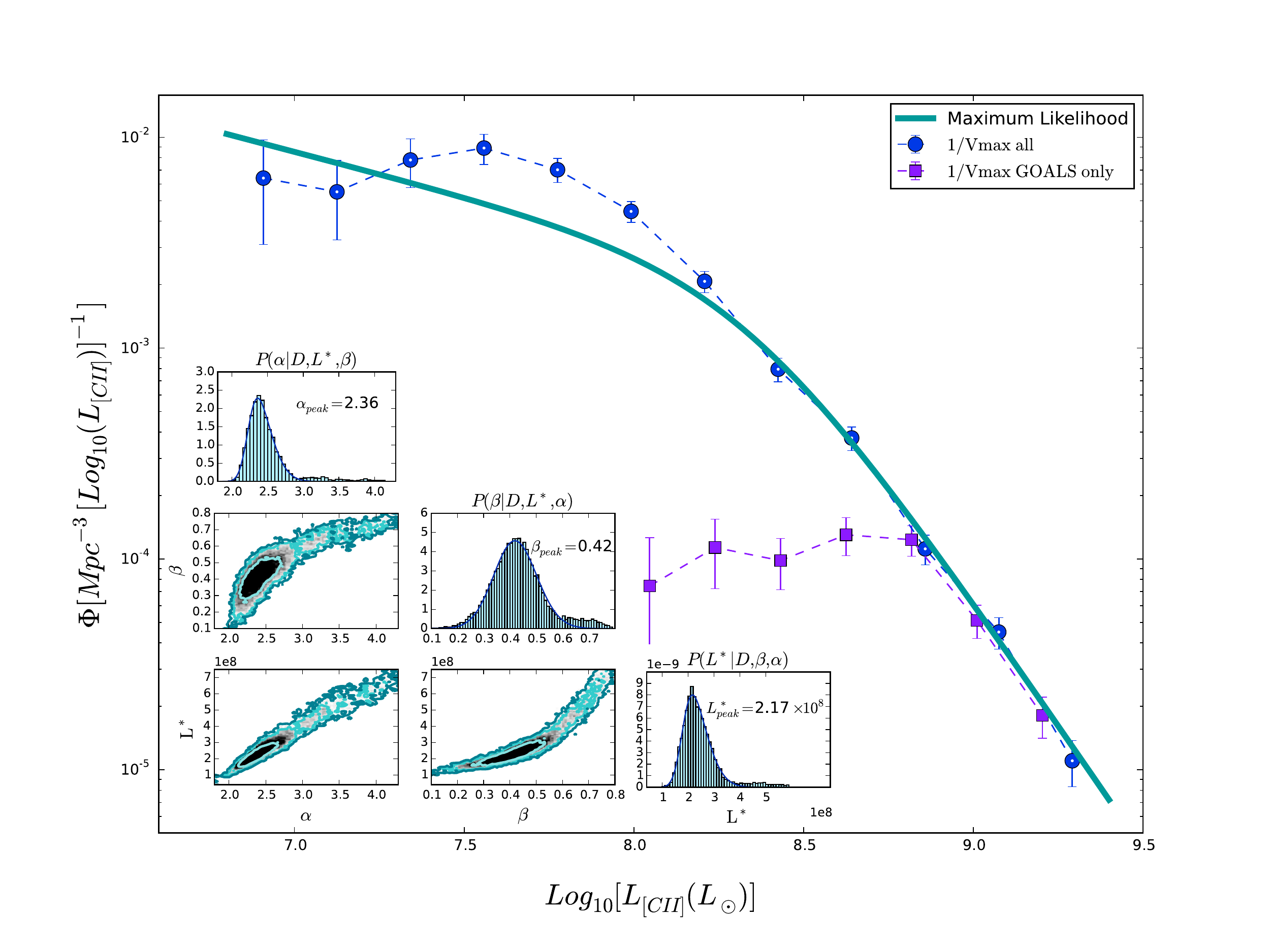}
\caption{The \CII\ line luminosity function. Blue circles are measured from the
1/V$_{max}$ method including all galaxies in the sample. Purple
squares are measured from the 1/V$_{max}$ using only the GOALS
galaxies with no extra correction. The cyan solid line is our estimate
from the MCMC code based on the modified STY maximum likelihood
estimator. The inset on the left corner show the 2d and 1d posterior distribution of $\alpha$,
$\beta$, and $L^{*}$. One, two and three sigma confidence contours are
plotted on the 2d distributions with solid blue lines.  }
\label{fig:LF}
\end{figure*}

 Minimum redshift ($z_{min}$) for all the galaxies in the sample is set by the cut
on the luminosity distance ($D_L > 1$ Mpc) as mentioned in the previous
section. The maximum redshift ($z_{max}$) a galaxy can have and still be included
in the sample is the maximum between the galaxy's actual redshift and a
redshift measured by comparing the \CII\ line flux ($\rm F_{[CII]}$) to the limiting
line flux ($\rm Flim_{[CII]}$) of the sample:

\begin{equation}
z_{max}=max( z ,0.5(\sqrt{1+4z(1+z)\sqrt{\frac{F_{[CII]}}{Flim_{[CII]}}}}-1)) .
\label{eq:zmax}
\end{equation} 

All cosmological calculations, such as co-moving distance and volume,
are simplified for galaxies with small redshifts. Here we adopt the
following equations: luminosity distance D$_{L}$\,=\,$z(1+z){c
  \over H_0}$; co-moving volume V$_c$\,=\,${4\pi \over 3}$D$^3_M$,
with D$_M$ being proper distance (=\,${cz \over H_0}$)
\citep{Hogg1999}. The co-moving maximum volume for each galaxy in the
sample is then calculated as:

\begin{equation}
 \rm V_{max}={4\pi \over 3}(D^3_M(z_{max}) - D^3_M(z_{min})).
\label{eq:vmax1}
\end{equation}

Finally, the luminosity function is the sum of (1/V$_{max,i}$) over
all galaxies, divided by the luminosity interval of $Log(L_{[CII]})
=0.2 L_{\odot}$ with which the luminosity function is binned. 

\begin{equation}
\phi( L) = {1 \over \Delta L} {\sum_{i}} {1 \over V_{max,i}} 
\label{eq:vmax}
\end{equation}

\begin{table}
\centering
\caption{1/V$_{max}$ Determination of the local \CII\ Luminosity
  Funcition}
\begin{tabular}{*{5}{c}}
\hline
\hline
$\rm Log_{10}(L_{[CII]} (L_{\odot}))$&$\Phi [Mpc^{-3}[Log_{10}(L_{[CII]})]^{-1}]$\\
\hline
6.9 & $0.00598\pm0.003456$ \\ 
7.1 & $0.01119\pm0.003633$ \\ 
7.3 & $0.00673\pm0.0019$ \\ 
7.5 & $0.00804\pm0.001463$ \\ 
7.7 & $0.00764\pm0.001041$ \\ 
7.9 & $0.00532\pm0.000599$ \\ 
8.2 & $0.00225\pm0.000276$ \\ 
8.4 & $0.00117\pm0.000138$ \\ 
8.6 & $0.00044\pm5.8e-05$ \\ 
8.8 & $0.00016\pm2.3e-05$ \\ 
9.0 & $6e-05\pm1e-05$ \\ 
9.2 & $2e-05\pm4e-06$ \\ 
\hline
\end{tabular}
\label{tbl:LF_vals}
\end{table}

\subsection{Maximum Likelihood Estimator}

The maximum likelihood estimator is a powerful tool in statistics
which estimates the parameters of a model given data. In observational
cosmology, it was first used by \citet*{Sandage1979} (hereafter STY) for deriving
luminosity functions of different types of galaxies. As mentioned
earlier, this method assumes a functional form for the luminosity function which elliminates
the need for binning the data. The most common model used in UV and optical studies is the Schechter function. This
is expected as the model was originally derived from a stellar mass
function (\citealt{Press1974}, \citealt{Schechter1976}). In infrared
however, more galaxies have been found in the bright end making a double power-law a better fit to
the luminosity function (e.g. \citealt{Soifer1987},
\citealt{Patel2013}). For our analysis, we tested both functional forms
and found that the double power-law is a much better fit to the shape
of the luminosity function, defined as:

\begin{equation}
\phi(L)=\phi^{*}((L/L^{*})^{\alpha}+(L/L^{*})^{\beta})^{-1}
\label{eq:double_power}
\end{equation}

where, $\phi^{*}$ is the normalization factor, $L^{*}$ is the
characteristic luminosity, and $\alpha$ and $\beta$ are the slopes of
the faint and bright end of the luminosity function. To find the
optimized parameters of the double-power law luminosity function given
our \CII\ line luminosities and uncertainties, we have to maximize the product ($\mathcal{L}$) of probabilities of
finding each galaxy in the sample:

\begin{equation}
\mathcal{L}=\prod_{i}P(L_{i}| \alpha , \beta , L^{*})
\label{eq:ML}
\end{equation}

where the probability of each galaxy in the sample is defined as:

\begin{equation}
P(L_{i}| \alpha , \beta , L^{*})=[\frac{\int_{L_{min}}^{\infty} \phi(L^{'})\times F(L^{'};L_{i},\sigma_{i})dL^{'}}{\int_{L_{min}}^{\infty}\phi(L)dL}]^{w_{i}}.
\label{eq:ML_probability}
\end{equation}

In measuring the probability for each galaxy the normalization factor
$\phi^{*}$ cancels out and at each luminosity ($L_{i}$) this
probability can be measured by the three parameters $\alpha$, $\beta$
and $L^{*}$ . Equation~\ref{eq:ML_probability} is a modified version
of what was used in the original STY. Here we account for both incompleteness in the sample and line uncertainties
of each galaxy. We account for incompleteness in the sample by
including weights ($w_{i}$) into Equation
\ref{eq:ML_probability}. These weights are the inverse of the completeness at each luminosity with the
simple idea that galaxies with same luminosity at the same distance
should have the same probability of being found. This is similar to
incompleteness corrections in spectroscopic samples (see for example
\citealt{Zucca1994}). The uncertainties in \CII\ line luminosity
estimates are taken into account by adding the error function ($F(L^{'};L_{i},\sigma_{i})$) to the probabilities. This is done by
assuming a normal distribution for luminosities with 1$\sigma$ errors as the
width of the distribution and summing over all luminosities. While
this is negligible for \herschel\ detected GOALS sources which have
small error bars, it is essential for the rest of the sample. This
factor is in essence similar to what was introduced in
\citet{Chen2003} for accounting for photometric redshift errors.

In practice, we need to vary parameters L$^{*}$, $\alpha$, and $\beta$
to find the maximum $\mathcal{L}$ as well as the posterior
distribution for each of the parameters. However, this is very
computationally expensive for a large enough grid with multiple
integration, three free parameters and over 600 galaxies. Therefore, rather than
measuring the probabilities over the whole grid, we perform a random
walk Markov Chain Monte Carlo (MCMC) to derive the desired parameters
and their uncertainties. To further speed up the calculations, we go
to the logarithm space and use the summation of probabilities rather
than multiplication to measure $\mathcal{L}$.

We run our MCMC program with hundred thousand steps randomly chosen
from the three dimensional grid (L$^{*}$, $\alpha$, and $\beta$). The
step size in each parameter is not fixed and is drawn from normal
distributions. We start by measuring the probabilities as described by Equation
\ref{eq:ML_probability} with an initial guess for the
parameters. These initial guesses do not need
to be precise as the first $5-10\%$ of steps are thrown away
(burn in process) and as long as the order of magnitude is correct the
process will converge to the optimized values. At each step, with the jump
to the new parameters we calculate the new $\mathcal{L}$ and compare it to
the previous one. In addition, a random acceptance rate scheme is
adopted, where the new parameters are accepted and added to the chain if the new
likelihood is either larger than the previous one or if their ratio is larger
than a random number drawn from a normal distribution. We choose the
jump size (width of the normal distributions) to get an acceptance
rate of $23\%$, which is shown theoretically to be the optimal
  value for an N-dimentional distribution (\citealt{roberts1997}).

\subsection{Results}

Figure \ref{fig:LF} represents our derived \CII\ line luminosity
function from both 1/V$_{max}$ (blue circles) and maximum likelihood
estimator (solid cyan line) methods as well as the posterior distributions of $\alpha$, $\beta$, and
$L^{*}$. We also show on Figure \ref{fig:LF} estimated \CII\ LF based on the
GOALS measurements only without any completeness correction (purple
squares) using the 1/V$_{max}$ method. It can be seen that while there
is good agreement at the very bright end, the LF stars to be very
incomplete and drops at luminosities below $log10(L_{[CII]}) \sim 8.7$. This clarifies the
importance of adding the estimated \CII\  fluxes from the rest of
the RBGS sample. We note here that the faintest and brightest \CII\ emitters in the sample are from the GOALS
galaxies with \herschel\  PACS observations and by including the rest
of the RBGS sample  we did not extrapolate to fainter or brighter \CII\ 
luminosities.

The maximum likelihood curve (solid cyan line) is calculated from
Skewed Gaussian function fits to the posterior distributions with best
fitted parameters shown in Table \ref{tbl:LF_param}. As can be seen
from the figure, there is a good agreement within uncertainties
between the maximum likelihood estimator and the 1/V$_{max}$
methods. In the maximum likelihood estimator method we accounted for
sources that might be missing using our derived completeness function,
in which we assigned random far-IR color to hundreds of galaxies at each
\CII\ luminosity and distance to determine whether they will be
detected by our survey. We note that we have drawn the random far-IR colors from a uniform
distribution in the color range of our sample due to lack of prior knowledge of
the true distribution of sources that we might be missing. The
agreement between the two methods of measuring the luminosity function
demonstrates the validity of this assumption.

Our completeness simulation is designed to account for potential \CII\
emitters that might be missed due to having $\rm S_{60\mu m}$ less
than the flux limit of the RBGS sample. However, we have assumed
similar far-IR color and dust temperature properties for the galaxies in this
simulation to those in the RBGS sample. An important question is, whether there exist galaxies with vastly
different properties that could change our results. Many recent
studies have shown the important role of dwarf galaxies in
understanding how galaxies form and evolve in general (e.g. \citealt{Dekel1986}, \citealt{Ferguson1994},
\citealt{Walter2007}, \citealt{Tolstoy2009}). More specifically, studies of optical
and UV luminosity functions found that dwarf galaxies can contribute
significantly to the faint end of the luminosity function
(e.g. \citealt{Liu2008}, \citealt{Alavi2016}). \cite{Madden2013} using
the \herschel\ Space Observatory provides a rich far-IR and submm photometric
and spectroscopic database for local dwarf galaxies covering a large
range in metallicity (Dwarf Galaxy Survey - DGS). Focusing on the
\herschel\  PACS spectroscopic data of DGS, \cite{Cormier2015}
found an increasing trend of \CII\ luminosity with increased
metallicity among the dwarfs, where they cover a large range of \CII\
luminosites ($10^{4-9} L_{\odot}$) and metallicities ($1/50$ - 1
$Z_{\odot}$). The low metallicity dwarfs which have low \CII\
luminosities have different dust temperature and far-IR colors
compared to normal local galaxies (see \citealt{Remyruyer2015}). As we
are only constraining the \CII\ luminosity function down to $\sim
10^{7} L_{\odot}$, the dwarf population in that \CII\ luminosity
regime will not have low metallicities ($12+log(O/H)> \sim
8.0$). These galaxies have similar far-IR colors and dust temperatures as the galaxies in our sample and therefore are
taken care of in the completeness measurement. However, low metallicity dwarf galaxies need to be taken into account if the \CII\
luminosity function extends to very faint \CII\ luminosities and they
can play an important role in constraining the very faint end of the
\CII\ luminosity function.

\begin{figure*}[]
\includegraphics[trim=0cm 0cm 0cm 0cm, clip, width=0.99\textwidth]{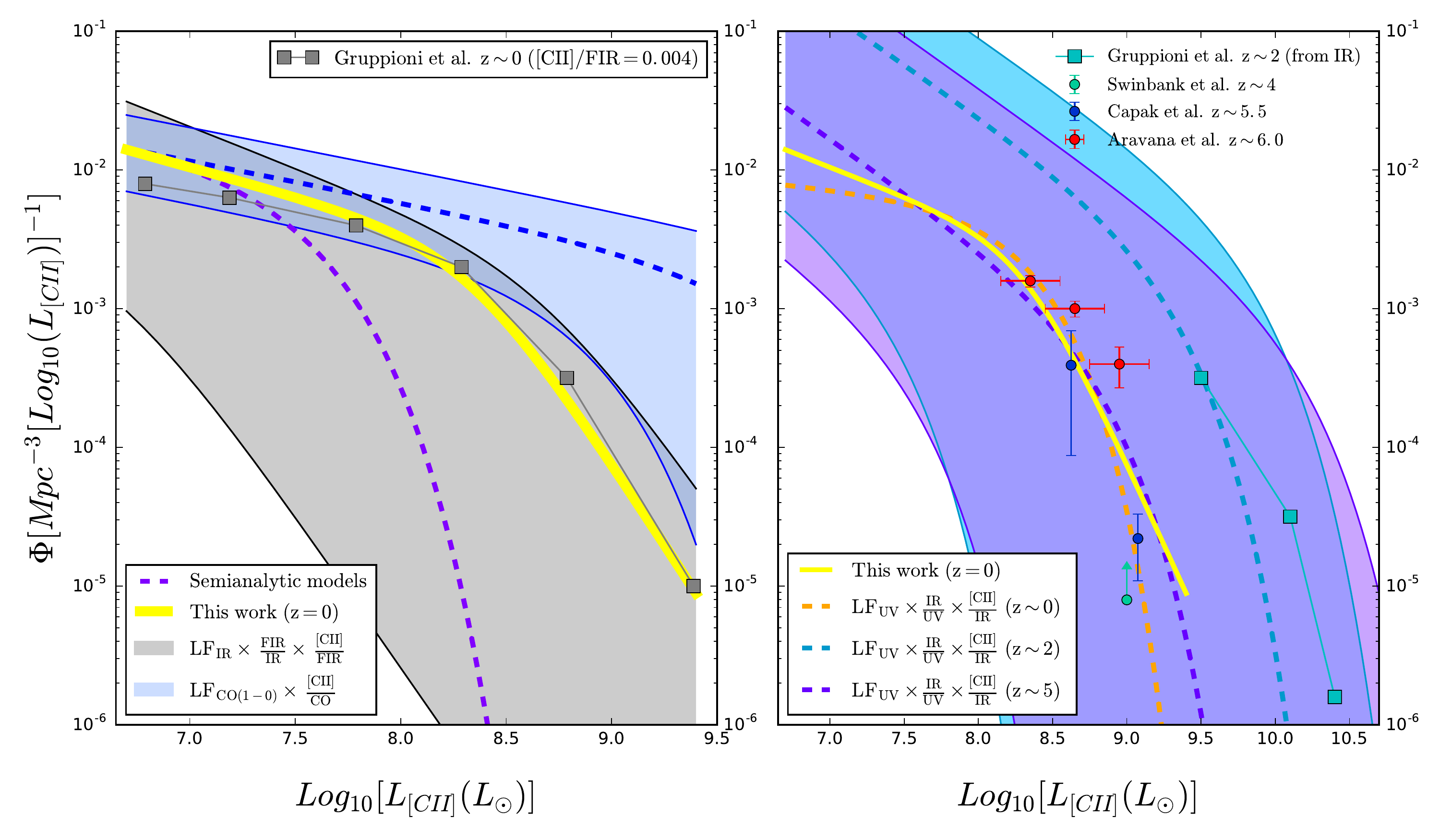}
\caption{Comparison of the derived \CII\ line luminosity
  function (solid yellow line) with other indirect estimates. Left: Conversion of IR luminosity function
  (\citet{Sanders2003}) to \CII\ line luminosity function assuming a range of [CII]/FIR shown with gray shaded
  region. \cite{Gruppioni2013} local measurements of IR luminosity
  function are plotted for comparison (gray squares) assuming a fixed
  ratio of 0.004 in \CII/FIR. Conversion of CO(1-0) luminosity function (Keres et al. 2003) to
  \CII\ LF is shown as light blue shaded region with the range adopted
  from \cite{Stacey1991} and blue dashed line represents a fixed value
  of $log([CII]/CO)=3.8$. Purple dashed
line shows the \CII\ luminosity function prediction from Popping et
al. (2016) based on semi-analytic models and radiative transfer codes. Right:
Prediction of evolution with redshift. Orange dashed line shows the
$z\sim 0$, blue dashed line and shaded
region represent $z\sim2$ and purple dashed line and shaded region
show $z\sim5$ based on UV observations with UV luminosity functions
adopted from Wyder et al. (2005), Alavi et al. (2014), Bouwens et
al. (2015), respectively. Green limit is from \cite{Swinbank2012} at
$z\sim 4$ and dark blue data points are estimates from \cite{Capak2015} at
$z\sim 5$. Also shown on the plot is the \CII\ luminosity function
estimate based on Gruppioni et al. (2013) $z\sim2$ IR luminosity
function (cyan squares).}
\label{fig:LFdisc}
\end{figure*}

\section{Discussion \label{sec:disc}}

Here, we compare our derived luminosity function to different local estimates obtained from other
indirect diagnostics in the left panel of Figure \ref{fig:LFdisc}. 

First, we compare our derived \CII\ luminosity function to an estimate
based on the IR luminosity function and assuming fixed \CII/FIR
ratios. For this comparison, we use the IR luminosity function of \cite {Sanders2003} derived from the
RBGS sample. We also take into account an extra factor to convert the
total IR luminosity function to far-IR luminosity function
($L_{IR}/L_{FIR} \sim 1.3$, \citealt{Chapin2009}). We note
here that the IR luminosity function from \cite{Chapin2009} matches
exactly the total IR luminosity function from \cite{Sanders2003}. The gray shaded region in the left
panel of Figure \ref{fig:LFdisc}, corresponds to the range of
\CII/FIR$=[0.0002-0.02]$ based on the scatter from
\cite{Santos2013}. We also show data points from the PEP/HerMES (\citealt{Gruppioni2013})
local IR luminosity function assuming a fixed ratio of \CII/FIR =0.004
with gray squares, which is chosen to match the \CII\ luminosity function. While fixing the \CII/FIR ratio to a single value
can result in good agreement between the IR and \CII\ luminosity
function, it is well known that not all galaxies can be described with
a single value of \CII/FIR. As previously shown in the literature
  (e.g. \citealt{Santos2013},\cite{Santos2014},\cite{Lutz2016}), FIR
  color and FIR surface brightness are among the most important
  observables linked to the variation of \CII/FIR, where the large variation is mostly among
galaxies with hotter far-IR SEDs (See Figure\ref{fig:sample}). We
note that converting existing IR LF into \CII\ LF using a constant ratio has serious flaws because
\CII/FIR ratio is not a constant number. For instance, low luminosity
galaxies could have stronger \CII\ emission, whereas IR brighter
ULIRGs/QSOs may have less. This could cause the shape of \CII\ LF
significantly differ from that of FIR LF.  We caution that such a
simple LF conversion is extremely crude, as illustrated by the large
shaded gray area.

Besides the \CII\ atomic fine structure line, the rotational
transitions of common interstellar molecules, predominantly carbon monoxide (CO), has
been used in the literature extensively to study the cool gas content
of galaxies.  While the \CII\ line is much stronger than
the CO(1-0) molecular line, an apparent linear correlation between
\CII\ and CO(1-0) intensity is reported in galactic star forming
regions as well as starburst extragalactic sources (e.g. \citealt{Crawford1985, Wolfire1989, Stacey1991}). We take the CO(1-0)
local luminosity function of \cite{Keres2003} and convert it to an
estimate of \CII\ luminosity function shown on the left panel of
Figure \ref{fig:LFdisc} with light blue color. Here, we assumed a range of
$log([CII]/CO)=2.5-4.5$ based on the lowest and highest values
presented in \cite{Stacey1991} for the galaxies NGC660 and LMC30Dor
respectively (similar range is observed by \cite{Hailey2010}). \cite{Stacey1991} showed that more active normal
starburst galaxies have $\rm L_{[CII]}/L_{CO}$ of $\sim$4000, while
more quiescent spiral galaxies have a factor of 2 lower ratios. The dashed blue line is derived if a fixed value of
$log([CII]/CO)=3.8$ corresponding to the average ratio reported by
\cite{Stacey1991} is assumed. \cite{Madden2000} also measured the $\rm
L_{[CII]}/L_{CO}$ for local low luminosity dwarfs and showed that the
ratio can get as high as $\sim$ 80000. As can be seen from the figure,
assuming the fixed average ratio reported by \cite{Stacey1991}, yields a good
agreement between the two luminosity functions at faint
ends but the discrepancy gets larger as one moves to the brighter
\CII\ luminosities. While some difference can be partly due to the
completeness correction applied in deriving the CO(1-0) luminosity
function, as well as forcing a Schechter functional form fit, the
large uncertainty is from the ratio of the lines. A better agreement between the two can be achieved
if a CO dependent ratio of $log([CII]/CO)$ is applied to the CO
luminosity function where galaxies with higher CO luminosity have
lower $[CII]/CO$ ratio compared to those with lower CO
luminosity. Theoretically this might be explained by CO being photodissociated
into C and $\rm C^{+}$ in low dust and metallicity environments by strong FUV field
from young stars (e.g. \citealt{Wolfire2010}, \citealt{Madden2013}).

In Figure \ref{fig:LFdisc} (left panel) we also compare our luminosity function measurement to the
recent estimate of \cite{Popping2016} using semi-analytic models (SAMs) and radiative
transfer models (magenta dashed line). \cite{Popping2016} studied the
evolution of both CO and \CII\ luminosity function from
$z=0-6$. However, their models under-predict the local \CII\ luminosity of FIR-bright galaxies
(Figure 3 of \cite{Popping2016}) which explains the very large disagreement in
the bright end seen here. \cite{Gruppioni2015} compared SFR function
derived from IR luminosity with those derived from four different
SAMs and found a similar trend at higher redshifts ($z\sim
2$). Similar disagreement has also been reported between the SAMs and the bright end of the CO luminosity function by
\cite{Vallini2016}. There, they suggest that the SAMs difficulty in modeling
the AGN feedback that affects the inflow/outflow of gas in the largest
and most massive galaxies, might explain the reason for this
difference. Nevertheless of the shape of the luminosity function at
$z=0$, their models predict that the number density of \CII\ line
emitting galaxies increases from $z=6$ to $z=4$, remains relatively
constant till $z=1$ and rapidly decreases towards $z=0$.

\subsection{Redshift Evolution}

Understanding the precise evolution of the \CII\ luminosity function would
require a larger sample of high redshift galaxies than already
exists. ALMA with full capability is ideal for acquiring
such statistical sample. There however exists limited observations and
limited high redshift \CII\ detections (e.g. \citet{Swinbank2012},
\citet{Capak2015}, \cite{Matsuda2015}). We show on the right panel of Figure
\ref{fig:LFdisc} where these measurements sit as well as a very rough
estimate of the \CII\ luminosity function at high redshifts based on
the UV and IR observations. 

The cyan circle on the right panel of Figure \ref{fig:LFdisc} represent
a lower limit from the ALMA detection of \CII\ in two $z\sim4$
galaxies by \cite{Swinbank2012} . In that study, they suggested a dramatic increase
from z=0 to z=4 in the number density at the bright end of the luminosity function in
contrast to what we see here. This is solely due to their lower number density
estimate at the bright end of \CII\ luminosity function at $z=0$.
  To estimate the $z=0$ luminosity function \cite{Swinbank2012} used
  \cite{Sanders2003} FIR luminosity function and the \CII/FIR with
 FIR luminosity corelation of \cite{Brauher2008}. To test the
  reliability of their method they also used \CII\ luminosities of 227
  galaxies compiled by \cite{Brauher2008}. As the data comes from a complex mix of observations, completeness
  measurement becomes a big issue.

 Using ALMA cycle 1 archival data (in band 7), \cite{Matsuda2015}
looked for \CII\ emission in $z\sim4$ galaxies and found no
significant emission. They presented upper limits to the
$z=4$ \CII\ luminosity function which is at least two orders of
magnitude larger than the \CII\ luminosity function expected from the
UV luminosity function. \cite{Capak2015} observed 9 $z\sim5$ normal
($\sim 1-4 L_{*}$) star forming galaxies using ALMA and detected \CII\
in all of the galaxies. They reported enhancement in the \CII\
emission relative to the FIR continuum
 and therefore a strong evolution in the interstellar medium
 properties in the very early universe. Blue circles on the right panel of Figure
\ref{fig:LFdisc} represent a very rough estimate of where these measurements sit
compared to the local luminosity function. To do this, we measure the volume for each observation using the area and the
redshift width of each ALMA pointing and as this was a targeted
observation of Lyman break galaxies we correct the volume using the number density of the Lyman
break galaxies (\citealt{Bouwens2015}).  Aside from these factors and
the low number statistics which makes these estimates very sensitive on choice of bins and therefore uncertain, it should be noted that
there might exist classes of galaxies that are faint in the UV and optical and therefore not
selected as LBGs at high redshifts that are bright in FIR and can
contribute to the luminosity function of \CII\ line. Recently,
  \cite{Aravana2016} identified fourteen \CII\ line emitting
  candidates using ALMA observations of optical dropout galaxies in
  the Hubble Ultra-Deep Field in the range $\rm 6<z<8$. Their data
  points are overplotted on the right panel of Figure \ref{fig:LFdisc}.

Also shown on the right panel of Figure \ref{fig:LFdisc} are estimates of \CII\ luminosity
function from UV observations at three different redshifts ($\rm z=0,\
2,\  \& 5$). As a crude estimate, we start with the UV luminosity
function, we adopt the values and uncertainties from \cite{Wyder2005}
for local galaxies, from \cite{Alavi2014} at $z=2$ and from \cite{Bouwens2015} at
$z=5$. To convert this to an IR estimate we use the IRX-$\beta$
relation (\citealt{Meurer1999}), which states that the ratio of dust
emission in the IR to UV emission (IRX) correlates with the UV
spectral slope $\beta$. We picked the ratio from the literature
(\citealt{Takeuchi2012}, \citealt{Alavi2014}, \citealt{Capak2015}) at each redshift based
on the reported average $\beta$ value and the best corresponding
IRX-$\beta$ curve (i.e. Calzetti like dust (\citealt{Calzetti2000}) for $z=0,2$ and SMC like
dust  (\citealt{Gordon2003}) for $z=5$). These correspond to $Log(L_{IR}/L_{1600})$ of 1.0,
0.7, and -0.2 for $\rm z=0,\ 2,\  \& 5$ respectively. The final factor
is an assumption for the $[CII]/IR$ ratio which yields an estimate of
the \CII\ luminosity function. Again this average ratio and its range is taken from the
literature to be 0.01 $\rm [0.001-0.03]$ at $\rm z=2$
(\citealt{Stacey2010}) and 0.01 $\rm [0.003-0.03]$ at $\rm z=5$
(\citealt{Capak2015}). At $z=0$ (orange dashed line) we only show the average curve to compare with our derived local \CII\
luminosity function. Overall the agreement between the UV
estimate and the actuall derived \CII\ luminosity function is good
despite all the assumptions that went into the estimate from the
UV. At $z=2$ ($z=5$), the cyan (purple) dashed line shows the estimated \CII\
luminosity function with the median assumptions and the cyan (purple) shaded
region corresponds to the whole possible range based on errors of the
UV luminosity function, the $IR/UV$ and the $[CII]/IR$. As can be seen from the
 width of the shaded regions, these are very uncertain estimates and
 ALMA observations are needed to constrain the picture. However,
 assuming the median values (dashed lines) we see a similar
 evolutionary trend as predicted by \cite{Popping2016} simulations,
 where the high redshift and local estimates are similar and the rise
 in the number densities is seen at intermediate redshifts ($z\sim 2$). 

 IR luminosity functions can exclude the uncertainy from
the IRX-$\beta$, but unfortunatly IR luminosity functions at high
redshifts ($z>3$) only overlap with the brightest part of our \CII\ luminosity
function. For example, \citet{Gruppioni2013} using \herschel\ PACS
selected galaxies estimated the IR luminosity function out to
$z=4$, but their highest redshift bins only cover \CII\
luminosities outside the range of our study. However, we use their IR
luminosity function at $z=0$ and $z=2$ and again assume a $[CII]/IR$
ratio (same as in our UV test above) and find perfect agreement for
the local measurement and agreement within the errors at  $z=2$. These
estimates of the \CII\ luminosity function are shown on Figure
\ref{fig:LFdisc} with gray squares on the left panel at $z=0$ and at $z=2$ with cyan
squares on the right panel.

In conclusion, we find tentative evidence which suggests that redshift
evolution of \CII\ LF may not be simply linear.  The volume density of
\CII\ emitters may increase significantly from $z=0$ to $z=2$, but at
$z=5-6$, \CII\ LF seems to return back to the similar level as $z=0$.
This redshift evolution behavior is similar to that of cosmic star formation
rate density which rose from early epochs to its peak value between
$z\sim$ 3 and 1 and drops towards the present time
(e.g. \citealt{Hopkins2006}, \citealt{Khostovan2015}). This evolution
of the cosmic star formation rate density is partly explained by the ISM masses
and the accretion/consumption of the molecular/total gas
(e.g. \citealt{Walter2014}, \citealt{Genzel2015}, \citealt{Scoville2016}).

\section{Summary}

In this paper we presented for the first time, the local \CII\ emission line
luminosity function using both \herschel\ PACS observed emission line data from the
GOALS survey as well as estimates based on the FIR emission for the rest
of the RBGS galaxies. This sample of 596 galaxies covers 91.3\% of the entire sky
(37,657 $deg^{2}$) and is complete at $S_{60\mu m} > \ 5.24\  Jy$. We
argue that in the absence of a blind deep \CII\ flux limited survey,
this is the best approach in estimating the local \CII\ luminosity
function. 

\begin{itemize}

\item Here, the luminosity function is estimated using both the $1/V_{max}$
as well as the STY maximum likelihood approach over the \CII\ luminosity range
$\sim 10^{7-9}$. The incompleteness function is measured over a grid
of \CII\ luminosity and distance by assigning hundreds of FIR
colors to each cell in the grid and recovering the $S_{60 \mu m}$ flux
density, to calculate the fraction of objects that could end up in the
final sample at each \CII\ luminosity and distance. We find that for the
majority of the sample the completeness is more than 80\%. We also showed that
low metallicity dwarf galaxies would not affect our \CII\ luminosity
function in the range of \CII\ luminosities covered by this work, but should be taken into
account if the luminosity function is to be extended to fainter \CII\
luminosities.

\item We compared our derived luminosity function with the FIR luminosity
functions from the literature. The $[CII]/FIR$ ratio is not a single value for
different galaxies and it varies with the average dust temperature covering the range $\sim 0.0002-0.02$. We
show that our derived \CII\ luminosity function lies in the range
defined by the $[CII]/FIR$ ratio and has a similar shape as the FIR
luminosity function.

\item The \CII\ luminosity function derived from the SAMs and
  radiative transfer models (\citealt{Popping2016}) deviates from our luminosity function, and the disagreement
gets larger at the bright end.

\item We also compared local CO(1-0) luminosity function of
  \cite{Keres2003}, to our \CII\ luminosity function assuming a range of
  $log([CII]/CO)=2.5-4.5$ from the literature and found that for the two luminosity
  functions to agree the $[CII]/CO$ value should be larger at fainter
  CO luminosities.  

\item ALMA with full capability will be ideal to acquire large samples of high
  redshift \CII\ emitters. While now there are only limited detections
  and therefore large uncertainty, we predict an evolution in
  the \CII\ luminosity function similar to that of the star formation
  rate density. We show that there are indications that the number density of \CII\
  emitters would increase from early times ($z\sim 5$) to its maximum
  value at $z\sim 2$ and decreases again to the present time. 

\end{itemize}

 We wish to thank the referee for helpful comments which improved the
content and presentation of this paper. SH wishes to thank Iary Davidzon and Hooshang Nayyeri for very useful
discussions. AF acknowledges support from the Swiss National Science Foundation.

\bibliography{ciibib.bib}

\end{document}